\documentstyle[11pt,aaspp4]{article}

\received{}
\accepted{}
\def\halpha{${\rm H} \alpha$}
\def\chisqnu{$\chi^2_\nu$}

\slugcomment{Submitted to ApJ.
Refereed version.
Draft: 981204}

\lefthead{Smith et al.}
\righthead{Keck Spectroscopy of GX 339--4}

\begin{document}

\title{Multiwavelength Observations of GX 339--4 in 1996.\\
III. Keck Spectroscopy}

\author{I. A. Smith\altaffilmark{1}, 
A. V. Filippenko\altaffilmark{2},
and D. C. Leonard\altaffilmark{2}}

\altaffiltext{1}{Department of Space Physics and Astronomy, 
Rice University, MS 108, 6100 South Main Street, Houston, TX 77005-1892}

\altaffiltext{2}{Department of Astronomy, University of California, Berkeley, 
CA 94720-3411}

\begin{abstract}

As part of our multiwavelength campaign of observations of GX 339--4
in 1996 we present our Keck spectroscopy performed on May 12 UT.
At this time, neither the ASM on the {\it Rossi X-Ray Timing Explorer} 
nor BATSE on the {\it Compton Gamma-Ray Observatory} detected the source.
The optical emission was still dominated by the 
accretion disk with $V \approx 17$ mag.
The dominant emission line is \halpha, and for the first time we are 
able to resolve a double peaked profile.
The peak separation $\Delta v = 370 \pm 40 ~{\rm km~s}^{-1}$.
Double peaked \halpha\ emission lines have been seen in the quiescent 
optical counterparts of many black hole X-ray novae.
However, we find that the peak separation is significantly smaller in 
GX 339--4, implying that the optical emission comes from a larger radius 
than in the novae.
The \halpha\ emission line may be more akin to the one in Cygnus X-1,
where it is very difficult to determine if the line is intrinsically double 
peaked because absorption and emission lines from the companion star dominate.

\end{abstract}


\section{Introduction}

Most Galactic black hole candidates exhibit at least two distinct spectral
states (see Liang \& Narayan 1997, Liang 1998, Poutanen 1998 for reviews).
In the hard state (= soft X-ray low state) the spectrum from $\sim$ keV 
to a few hundred keV is a hard power law (photon index $1.5 \pm 0.5$) with 
an exponential cutoff.
This can be interpreted as inverse Comptonization of soft photons.
In the soft state (often, but not always, accompanied by the soft X-ray 
high state), the spectrum above $\sim 10$ keV is a steep power law 
(photon index $> 2.2$) with no detectable cutoff out to $\sim$ MeV.
This multi-state behavior is seen in both persistent sources 
(e.g. Cygnus X-1) and transient black hole X-ray novae (BHXRN, 
e.g. GRS 1009--45).
GX 339--4 is unusual in that it is a persistent source, being detected by 
X-ray telescopes most of the time, but it also has nova-like flaring states.

In 1996, we performed a series of multiwavelength observations of GX 339--4
when it was in a very low state.
This paper is one of a series that describe the results of this campaign.
In Paper I (\cite{smi99I}) we discuss the radio, X-ray, and gamma-ray
daily light curves and spectra obtained in 1996 July.
In Paper II (\cite{smi99II}) we discuss the X-ray timing analysis from 
1996 July.
In B\"ottcher, Liang, \& Smith (1998) we use the GX 339--4 spectral data 
to test our detailed self-consistent accretion disk corona models.
These papers expand significantly on our preliminary analyses 
(\cite{smi97a,smi97b}).

In this paper we focus on Keck spectroscopy performed on 1996 May 12 UT.
Beginning with the discovery observations in the 1970s
(\cite{pen75,dox79,gri79})
it was clear that the optical counterpart to GX 339--4 is highly variable
(e.g. \cite{motch82,motch83,motch85,call92}).
For example, one month prior to our Keck observations, when the source
was in a similar low state, it was found to have 16 second optical
quasi-periodic oscillations (\cite{steim97}).

In previous spectroscopic observations, the optical emission has always been 
dominated by the highly variable accretion disk, preventing a direct 
observation of the spectrum of the companion star.
This is also true of our Keck observations.
For the first time we are able to resolve the \halpha\ emission line,
showing it is double peaked.
This is similar to the \halpha\ emission lines from the quiescent optical
counterparts of many BHXRN.
However, we show that the peak separation in GX 339--4 is much smaller
than in the BHXRN.
Instead, the \halpha\ emission line may be more akin to the one in Cygnus X-1.

\section{Observations and Reductions}

Our Keck observations were performed on 1996 May 12 UT (MJD 50215).
At this time, the ASM on the {\it Rossi X-Ray Timing Explorer} did not detect
the source, with count rates of
$-0.636 \pm 0.395$ cps in the 1.3 -- 3.0 keV band,
$0.027 \pm 0.331$ cps in the 3.0 -- 5.0 keV band, and
$0.420 \pm 0.403$ cps in the 5.0 -- 12.1 keV band.
Similarly, using the Earth occultation method, BATSE on the 
{\it Compton Gamma-Ray Observatory} also did not detect it with flux 
$-0.0024 \pm 0.0061 ~{\rm photons~cm}^{-2}{\rm s}^{-1}$ in the 
20 -- 100 keV band.
The source was very faint or not detected for weeks on either side of this
time: see Figure 1 of Smith et al. (1997a) for the RXTE light curve.
This is as close to ``quiescence'' as the high-energy emission comes in
GX 339--4.

Our observations of GX 339--4 were made during one of the same runs in 
which we studied the BHXRN Nova Ophiuchi 1977 in quiescence and found 
its orbital period and mass function; see Filippenko et al. (1997) and
Harlaftis et al. (1997) for full instrumental and reduction details.
The Low-Resolution Imaging Spectrometer was used at the Cassegrain focus
of the Keck I telescope.
Two consecutive 400 second exposures were taken, starting at 11:34 and
11:41 UT.
A 1\arcsec\ slit was used, but the seeing was poor (2.2\arcsec) and
variable, probably due to the high airmass (2.8) and possibly affected
by thin clouds.

\section{Results}

\subsection{Whole Spectrum}

The original goal of our observations was to study the source in optical
quiescence to determine the nature of the companion star; it was hoped that 
the lack of detectable high-energy emission would make this possible.
However, we found that the optical emission was still dominated by the 
accretion disk.
Figure 1 shows the two consecutive 400 s spectra.
The data are not fully photometric but give ${\rm V} \approx 17$ mag.
This is typical for the source in the low state, ${\rm V} \approx 15.4 - 20.2$:
here we include both the results reported for the traditional 
``soft X-ray low'' state (${\rm V} \approx 15.4 - 17$) and the sometimes 
discussed ``off'' state (${\rm V} \approx 17.7 - 20.2$), since it is now
generally believed that the off state is a lower luminosity low hard
state (\cite{vdk95}).

There is no artificial offset in the two curves in Figure 1.
The ratio of the spectra is consistent with being constant across the whole 
range, including the emission lines.
Such a variability and constant spectral shape agrees with previous 
observations.
For example, in the high state in 1986 June/July it was found that there 
were large changes in $V$, but no accompanying variations in the (\bv) index
(\cite{corbet87}).
However, since the slit was narrow relative to the seeing conditions at
the time of the observation, we can make
no accurate statements regarding the true variability of the source.

The Keck continua can be approximately fit by a linear relationship.
Extrapolating this, we find that the flux density goes to zero at
$\sim 4000$ \AA, easily consistent with the lack of observed X-ray emission
(5 keV corresponds to 2.45 \AA).
However, given the relatively narrow wavelength range of the Keck spectrum, 
and a misalignment between the slit position angle (60\arcdeg) and the
parallactic angle (175\arcdeg) which can distort the continuum shape at the
edges of the spectrum (\cite{fil82}),
we cannot make conclusive statements about the spectrum at higher energies.

\subsection{\halpha\ Lines}

The dominant emission line in our spectra is \halpha.
Its equivalent width (EW) of $\sim 6.5$ \AA\ is similar to the lines seen
in previous GX 339--4 observations (\cite{gri79,cow87}).
However, this is small compared to other BHXRN; for example, the \halpha\ line
in Nova Oph 1977 had EW $= 85$ \AA\ during the same observing run 
(\cite{fil97}).

An expanded view of the \halpha\ line in GX 339--4 is shown in Figure 2.
For the first time we are able to resolve a double peaked line about the 
rest wavelength of 6562.80 \AA\ in both spectra.  

We used a linear fit to the continuum around the line to determine the rms
noise.
Adding a single Gaussian line (with unconstrained centroid) does not give a 
good fit in either of our spectra.
For the upper spectrum, the reduced \chisqnu = 1.57 for $\nu = 51$ 
degrees of freedom.
The probability that a random set of data points would give a value of
\chisqnu\ as large or larger than this is $Q = 5.6 \times 10^{-3}$.
The lower spectrum gives very similar results: \chisqnu = 1.54, $\nu = 73$, 
$Q = 2.2 \times 10^{-3}$.

Using two unconstrained Gaussian lines greatly improves the fits.
Figure 2 shows the best fit results.
The fit to the upper spectrum now has \chisqnu = 1.23, $\nu = 48$, $Q = 0.13$,
while the lower spectrum now has \chisqnu = 0.95, $\nu = 70$, $Q = 0.59$.
It therefore appears that the \halpha\ line in GX 339--4 is double peaked, 
as is the case for the quiescent optical counterparts of many BHXRN.

The best two Gaussian fits have different fit parameters for the two spectra.
However, if we multiply the fit to the upper spectrum by a constant 
1/1.062 given by the ratio of the continua shown in Figure 1, we find
that this gives an adequate fit to the lower spectrum:
\chisqnu = 1.11, $Q = 0.25$.
Thus we are unable to claim there is significant variability in the 
\halpha\ lines in these data.
Searching for this variability is the subject of an ongoing project, at
which point it will also be possible to use more realistic emission-line
profiles (e.g. Orosz et al. 1994, and references therein).

The separation of the peaks is $\Delta \lambda = 8.0 \pm 0.8$ \AA,
implying a $\Delta v = 370 \pm 40 ~{\rm km~s}^{-1}$.
If we assume this peak emission comes from a circular Keplerian orbit,
it would be at a distance $4 \times 10^{11} (M/M_{\sun})$ cm, 
where $M$ is the mass of the black hole.
This would be in the outer regions of the accretion disk.
This distance can be contrasted with the suggestion by Motch et al. (1982) 
that the 20 second optical QPO they detected came from a ring orbiting with 
a Keplerian period of 20 seconds at a radius of $10^9$ cm. 

\subsection{\ion{He}{1} Lines}

The only other emission lines in the spectra are from \ion{He}{1}.
For $\lambda$5875, the combined equivalent width is $1.3 \pm 0.3$ \AA,
although this is difficult to fit accurately because of the adjacent NaD 
interstellar absorption lines.
For $\lambda$6678, the combined equivalent width is $1.0 \pm 0.1$ \AA. 

\subsection{\ion{Li}{1} Line}

We do not see any evidence for an absorption line from 
\ion{Li}{1} at $\lambda$6708 \AA.
This result agrees with previous observations of GX 339--4 (\cite{soo97}).

\section{Comparison with other Galactic Black Hole Candidates}

A double peaked \halpha\ emission line has been seen in several BHXRN
in quiescence, showing that there is optical emission from their accretion
disks long after the X-ray nova event.
Table 1 lists representative values for their peak separations.
The remarkable feature that has already been noted in previous
studies is that the peak separations in these different sources are all 
surprisingly similar.

The \halpha\ peak separation in GX 339--4 clearly shows that this source is 
different from the usual BHXRN.
This may not be too surprising.
Although GX 339--4 had no detectable high-energy emission around the time of
our Keck observation, the source did not exhibit a very long period of
``quiescence'' in the same way as the other BHXRN.
The narrower \halpha\ peak separation in GX 339--4 implies that the
optical emission comes from a larger Keplerian radius than in the BHXRN, 
which may be a clue to its different behavior.

GX 339--4 is most often compared to Cygnus X-1.
Unfortunately, for Cygnus X-1 the absorption and emission lines from the 
companion star dominate, and only the wings of the \halpha\ emission line are 
detected (\cite{can95,sow98}).
It is therefore very difficult to determine if the \halpha\ emission line
in Cygnus X-1 is double peaked.
However, the width of the base of the \halpha\ emission line in Cygnus X-1
is quite similar to that in GX 339--4, suggesting that the optical
emission from their accretion disks may be similar.

It is also interesting to note that the peak separation was
$\sim 350 - 550 ~{\rm km~s}^{-1}$ in the \halpha\ emission lines 
from GRO J1655--40 {\it during outbursts} (\cite{soria98}).
This is a superluminal jet source, and may indicate a connection
to the suggested radio jet in GX 339--4 (\cite{fen97}).

\acknowledgments

We thank Philippe Durouchoux for suggesting we check for a \ion{Li}{1} line.
We also thank the referee for carefully reading the manuscript and providing
useful suggestions and clarifications.
This work was supported by NASA grants NAG 5-1547 and 5-3824 at 
Rice University, and NSF grant AST-9417213 at UC Berkeley.
This work made use of the RXTE ASM data products provided by the ASM/RXTE teams
at MIT and at the RXTE SOF and GOF at NASA's Goddard Space Flight Center.
The BATSE daily light curves were provided by the Compton 
Observatory Science Support Center at NASA's Goddard Space Flight Center.

\clearpage

\begin{figure} \plotone{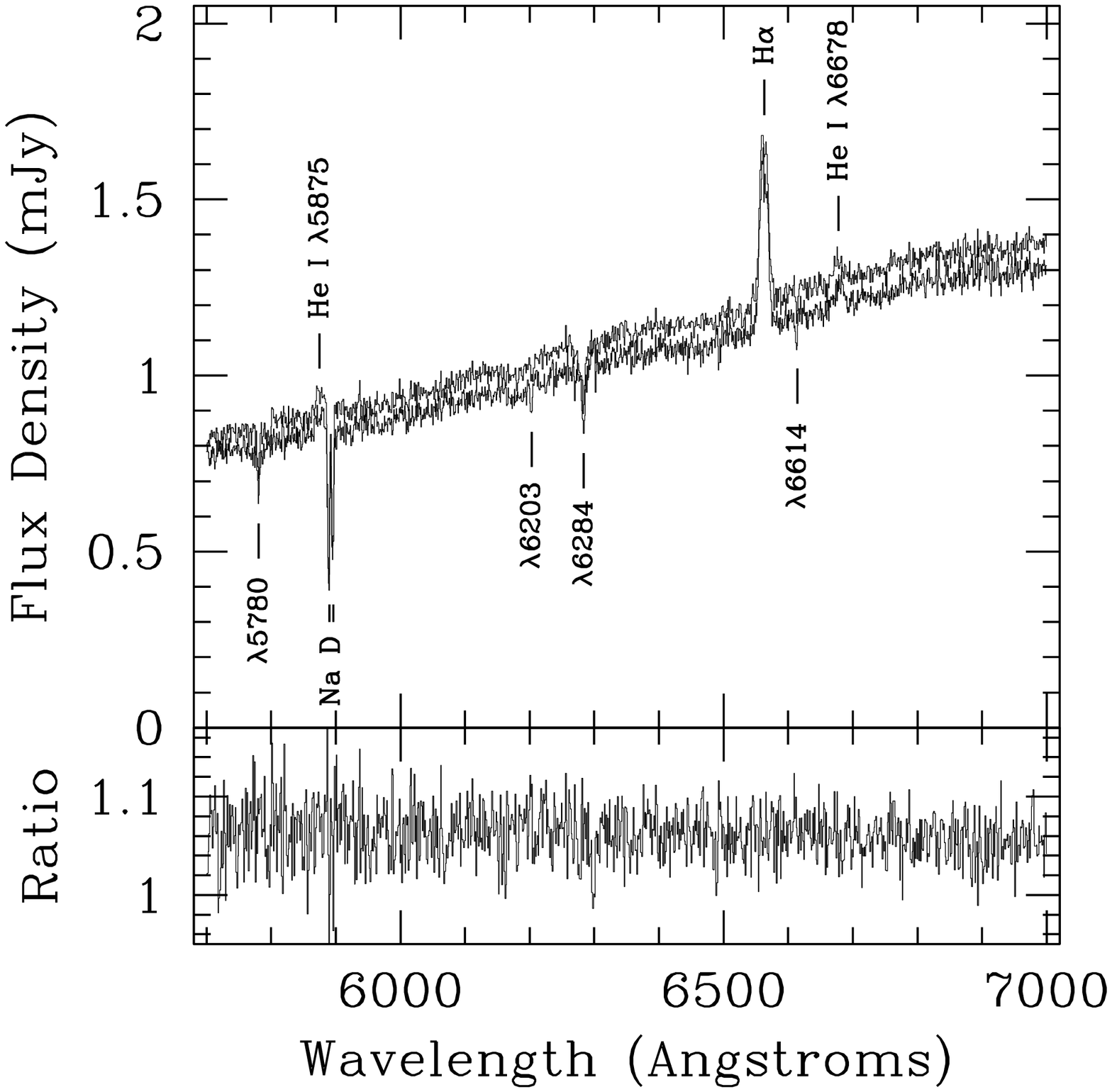} \begin{center}
\figcaption{1996 May 12 Keck observations of GX 339--4.
The offset between the two spectra may be due to intrinsic source variability
and/or variable observing conditions.
The absorption lines are all interstellar.
The ratio of the two spectra is fit by a constant = 1.062, with
\chisqnu = 0.97, $Q = 0.78$.}
\end{center} \end{figure}

\begin{figure} \plotone{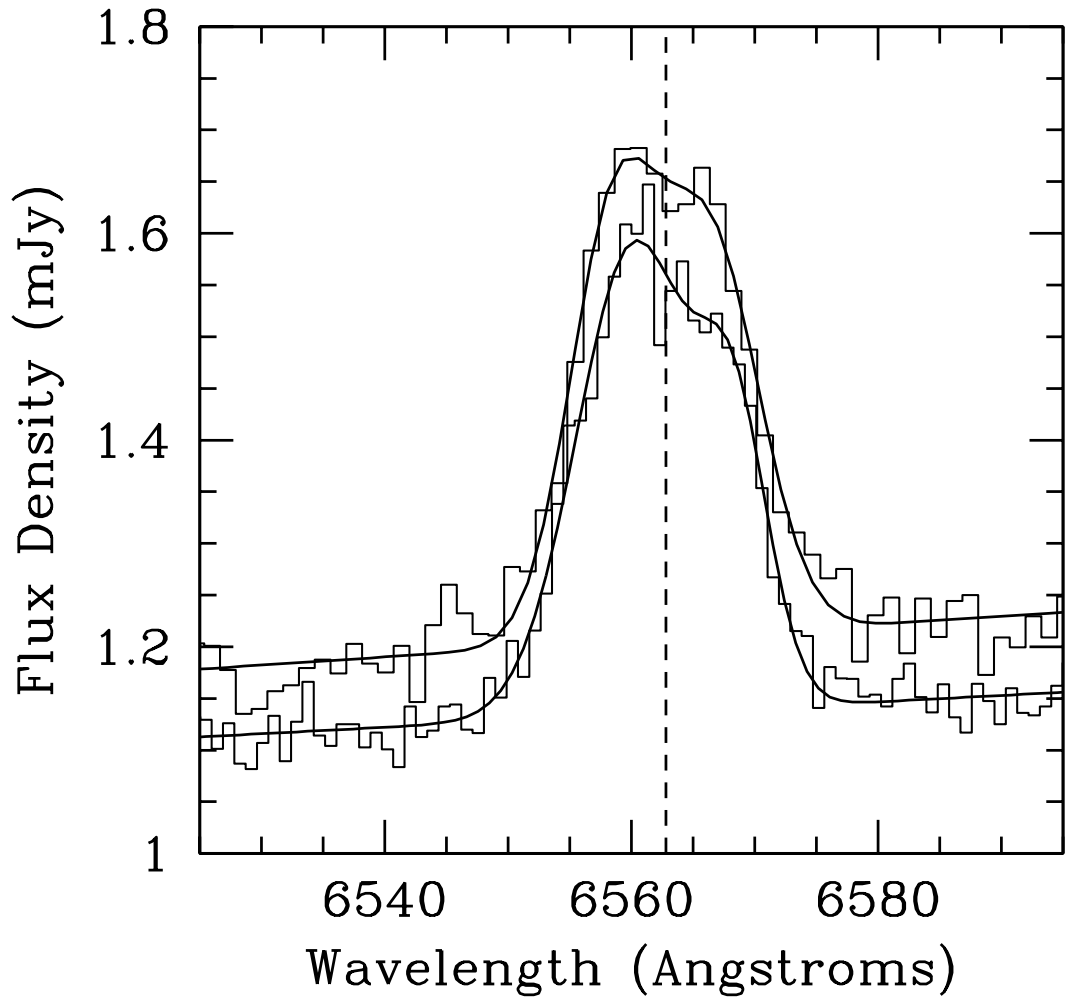} \begin{center}
\figcaption{1996 May 12 Keck observations of the \halpha\ emission in 
GX 339--4.
The dashed line shows the rest wavelength of 6562.80 \AA.
The solid curves show the best double-Gaussian plus linear continuum fits.}
\end{center} \end{figure}

\clearpage

\begin{deluxetable}{lrl}
%\small
%\footnotesize
%\scriptsize
\tablewidth{0pt}
\tablecaption{Typical peak separation $\Delta v$ (${\rm km~s}^{-1}$) of 
\halpha\ lines in galactic black hole candidates.}
\tablehead{
\colhead{Source} & \colhead{$\Delta v$} & 
\colhead{References}}
\startdata
GRO J0422+32 & 900 & Orosz \& Bailyn (1995), Filippenko, Matheson, \& Ho (1995)
 \nl
Nova Muscae 1991 & 900 & Orosz et al. (1994)\nl
Nova Oph 1977 & 900 & Remillard et al. (1996), Filippenko et al. (1997) \nl
GRS 1009--45 & 1000 & Shahbaz et al. (1996) \nl
A0620--00 & 1100 & Orosz et al. (1994)\nl
GS 2000+25 & 1400 & Casares et al. (1995), Filippenko, Matheson, \& Barth 
(1995) \nl
\nl
GX 339--4 & 370 & This paper \nl
\enddata
\end{deluxetable}


\clearpage

\begin{thebibliography}{}

\bibitem[B\"ottcher, Liang, \& Smith]{bls98}
B\"ottcher, M., Liang, E. P., \& Smith, I. A. 1998, \aap, 339, 87

\bibitem[Callanan et al. 1992]{call92}
Callanan, P. J., Charles, P. A., Honey, W. B., \& Thorstensen, J. R.
1992, \mnras, 259, 395

\bibitem[Canalizo et al. 1995]{can95}
Canalizo, G., Koenigsberger, G., Pena, D., \& Ruiz, E. 1995, 
Rev. Mex. Astron. Astrofis., 31, 63

\bibitem[Casares, Charles, \& Marsh]{cas95}
Casares, J., Charles, P. A., \& Marsh, T. R. 1995, \mnras, 277, L45

\bibitem[Corbet et al. 1987]{corbet87}
Corbet, R. H. D., Thorstensen, J. R., Charles, P. A., Honey, W. B.,
Smale, A. P., \& Menzies, J. W. 1987, \mnras, 227, 1055

\bibitem[Cowley, Crampton, \& Hutchings 1987]{cow87}
Cowley, A. P., Crampton, D., \& Hutchings, J. B. 1987, \aj, 92, 195

\bibitem[Doxsey et al. 1979]{dox79}
Doxsey, R., Grindlay, J., Griffiths, R., Bradt, H., Johnston, M.,
Leach, R., Schwartz, D., \& Schwartz, J. 1979, \apjl, 228, L67

\bibitem[Fender et al. 1997]{fen97}
Fender, R. P., Spencer, R. E., Newell, S. J., \& Tzioumis, A. K.
1997, \mnras, 286, L29

\bibitem[Filippenko 1982]{fil82}
Filippenko, A. V. 1982, \pasp, 94, 715

\bibitem[Filippenko, Matheson, \& Barth 1995]{fil95b}
Filippenko, A. V., Matheson, T., \& Barth, A. J. 1995, \apjl, 455, L139

\bibitem[Filippenko, Matheson, \& Ho 1995]{fil95a}
Filippenko, A. V., Matheson, T., \& Ho, L. C. 1995, \apj, 455, 614

\bibitem[Filippenko et al. 1997]{fil97}
Filippenko, A. V., Matheson, T., Leonard, D. C., Barth, A. J., 
\& Van Dyk, S. D. 1997, \pasp, 109, 461

\bibitem[Grindlay 1979]{gri79}
Grindlay, J. E. 1979, \apjl, 232, L33

\bibitem[Harlaftis et al. 1997]{har97}
Harlaftis, E. T., Steeghs, D., Horne, K., \& Filippenko, A. V. 1997, \aj,
114, 1170

\bibitem[Liang 1998]{lia98}
Liang, E. P. 1998, \physrep, 302, 67

\bibitem[Liang \& Narayan 1997]{lia97}
Liang, E., \& Narayan, R. 1997,
in Proceedings of the Fourth Compton Symposium Part One,
ed. C. D. Dermer, M. S. Strickman, \& J. D. Kurfess (New York: AIP), 461

\bibitem[Motch, Ilovaisky, \& Chevalier 1982]{motch82}
Motch, C., Ilovaisky, S. A., \& Chevalier, C. 1982, \aap, 109, L1

\bibitem[Motch et al. 1983]{motch83}
Motch, C., Ricketts, M. J., Page, C. G., Ilovaisky, S. A., \& Chevalier, C.
1983, \aap, 119, 171

\bibitem[Motch et al. 1985]{motch85}
Motch, C., Ilovaisky, S. A., Chevalier, C. \& Angebault, P. 1985, 
Space Sci. Rev., 40, 219

\bibitem[Orosz \& Bailyn 1995]{oro95}
Orosz, J. A., \& Bailyn, C. D. 1995, \apjl, 446, L59

\bibitem[Orosz et al. 1994]{oro94}
Orosz, J. A., Bailyn, C. D., Remillard, R. A., McClintock, J. E.,
\& Foltz, C. B. 1994, \apj, 436, 848

\bibitem[Penston et al. 1975]{pen75}
Penston, M. V., Penston, M. J., Murdin, P., \& Martin, W. L. 1975, \mnras,
172, 313

\bibitem[Poutanen 1998]{pou98}
Poutanen, J. 1998, in Theory of Black Hole Accretion Disks
(Cambridge: Cambridge U. Press), in press

\bibitem[Remillard et al. 1996]{rem96}
Remillard, R. A., Orosz, J. A., McClintock, J. E., \& Bailyn, C. D. 1996,
\apj, 459, 226

\bibitem[Shahbaz et al. 1996]{sha96}
Shahbaz, T., Van der Hooft, F., Charles, P. A., Casares, J., 
\& Van Paradijs, J. 1996, \mnras, 282, L47

\bibitem[Smith et al. 1997a]{smi97a}
Smith, I. A., et al. 1997a, 
in Proceedings of the Fourth Compton Symposium Part Two,
ed. C. D. Dermer, M. S. Strickman, \& J. D. Kurfess (New York: AIP), 932

\bibitem[Smith et al. 1997b]{smi97b}
Smith, I. A., et al. 1997b, 
poster at the HEAD of the AAS, Estes Park, CO

\bibitem[Smith et al. 1999]{smi99I}
Smith, I. A., et al. 1999, \apj, submitted

\bibitem[Smith \& Liang 1999]{smi99II}
Smith, I. A., \& Liang, E. P. 1999, \apj, submitted

\bibitem[Sood, James, \& Durouchoux 1997]{soo97}
Sood, R. K., James, S., \& Durouchoux, Ph. 
1997, in 2nd INTEGRAL Workshop: The Transparent Universe, ESA SP-382, 237

\bibitem[Soria et al. 1998]{soria98}
Soria, R., Wickramasinghe, D. T., Hunstead, R. W., \& Wu, K.
1998, \apj, 495, L95

\bibitem[Sowers et al. 1998]{sow98}
Sowers, J. W., Gies, D. R., Bagnuolo, W. G., Shafter, A. W., 
Wiemker, R., \& Wiggs, M. S. 1998, \apj, 506, 424

\bibitem[Steiman-Cameron et al. 1997]{steim97}
Steiman-Cameron, T. Y., Scargle, J. D., Imamura, J. N., \&
Middleditch, J. 1997, \apj, 487, 396

\bibitem[Van der Klis 1995]{vdk95}
Van der Klis, M. 1995, in X-Ray Binaries, 
ed. W. H. G. Lewin, J. van Paradijs, \& E. P. J. van den Heuvel
(Cambridge: Cambridge U. Press), 252

\end{thebibliography}
\end{document}